\documentclass{iopart}

\usepackage{iopams}

\newcommand{\realni}{\ensuremath{\mathbb{R}}}
\newcommand{\cG}{{\cal G}}
\newcommand{\cH}{{\cal H}}

\newcommand{\cM}{{\cal M}}
\newcommand{\fg}{{\mathfrak g}}
\newcommand{\fh}{{\mathfrak h}}
\newcommand{\diag}{{\mathop{\rm diag}\nolimits}}
\newcommand{\raz}{ \rule{0ex}{2.5ex} }
\newcommand{\del}{\partial}
\newcommand{\ds}{\displaystyle}
\newcommand{\lc}{\varepsilon}
\newcommand{\hodge}{\star}
\newcommand{\pb}[2]{\{\,  {#1} \,,\, {#2} \,\}  }
\newcommand{\db}[2]{{ \{\,  {#1} \,,\, {#2} \,\}_D} }

\begin{document}

\title{Hamiltonian analysis of the $BFCG$ theory for the Poincar\'e 2-group}

\author{A Mikovi\'c$^{1,2}$, M A Oliveira$^2$ and M Vojinovi\'c$^2$}

\address{$^1$ Departamento de Matem\'atica Universidade Lus\'ofona de Humanidades e Tecnologias, Av. do Campo Grande 376, 1749-024 Lisboa, Portugal}
\address{$^2$ Grupo de F\'isica Matem\'atica, Faculdade de Ci\^ encias da Universidade de Lisboa, Campo Grande, Edif\'icio C6, 1749-016  Lisboa, Portugal}
\eads{\mailto{amikovic@ulusofona.pt}, \mailto{masm.oliveira@gmail.com}, \mailto{vmarko@ipb.ac.rs}}

\begin{abstract}
We perform the full Hamiltonian analysis of the topological $BFCG$ action based on the Poincar\'e 2-group. The Hamiltonian of the theory is constructed, and the algebra of constraints is computed. The Dirac brackets are evaluated, and the second class constraints are then eliminated from the theory. The results are contrasted to those of the topological Poincar\'e gauge theory, which is equivalent to the $BFCG$ model at the level of the classical action, but has a very different Hamiltonian structure.
\end{abstract}

\submitto{\CQG}


\section{\label{SecIntroduction}Introduction}

The problem of quantization of gravitational field is one of the fundamental
problems in theoretical physics. One of the approaches to the construction of a theory
of quantum gravity is called Loop Quantum Gravity (LQG), which began with the
seminal work by Ashtekar \cite{AstekarPrviRad} and has since developed in a large research discipline \cite{RovelliBook}.
Research in LQG is focused on two complementary fronts. The approach used on one front is to perform the canonical quantization of the gravitational field, which is known as canonical LQG. This is done by rewriting General Relativity (GR) in the Hamiltonian form, by using connection variables and their momenta, with the goal of constructing the appropriate physical Hilbert space. This amounts to working with the the spin-network states. On the other front, the approach is to perform the path-integral quantization of GR by using the connection variables, and this requires using a spacetime triangulation where the triangles carry spin labels \cite{RovelliVidottoBook}. The resulting theories are called spin-foam models. 

The spin-foam path integrals can be viewed as transition amplitudes from one spin-network state to another one, via the corresponding spin foam. This picture provides a relationship between the canonical and the path-integral formulations. Although one can couple fermions in canonical LQG, it is impossible to couple correctly the fermions in the spin-foam formulation. This is a consequence of the fact that there are no edge lengths in a spin-foam model, which is related to the absence of the tetrads in the Plebanski action. Furthermore, the classical limit of a spin-foam model is the area-Regge theory \cite{mvea,Mikovic2013}, which is based on a twisted geometry, where metric is not always defined.

Recently a new idea has been proposed for the spin-foam approach, in the form of the spin-cube models \cite{Mikovic2013,TwoPoincare}. These models represent a categorical generalization of the spin-foam models. One can generalize the path integral construction based on a group to the construction based on a 2-group. For the basic notions of 2-group theory see \cite{BaezHuerta2011,BaezBaratinFreidelWise2012}. The construction of the spin-cube path integral rests on a reformulation of GR as a constrained $BFCG$ theory. The $BFCG$ theory is a categorical generalization of the well-known BF theory. The consequence for the path integral is that it becomes a sum over the labels for the edges, triangles and the tetrahedrons in a triangulation. Hence one obtains a sum of amplitudes for a colored 3-complex, therefore the name spin cube. This also implies that a spin-cube model can describe a transition amplitude from a spin-foam state to another spin-foam state, so that one would like to study a canonical quantization of the $BFCG$ formulation of GR in a spin-foam basis.

In particular, given a Lie group $\cG$ and its Lie algebra $\fg$, one can assign to it an action of the form
\begin{equation}
S_{BF} = \int_{\cM} \langle B \wedge F \rangle_{\fg}\,,
\end{equation}
where $B$ is a $\fg$-valued Lagrange multiplier two-form, while $F=\rmd A+A\wedge A$ is the curvature two-form for the $\fg$-valued connection one-form $A$. The $\langle\;\,,\;\rangle_{\fg}$ represents the invariant nondegenerate symmetric bilinear form in $\fg$. The $BF$ theory relevant for the construction of spin-foam models is based on the Lorentz group $SO(3,1)$. A categorical generalization of the $BF$ theory is based on the concept of a $2$-group, constructed of two groups $\cG$ and $\cH$ in a particular way (see \cite{BaezHuerta2011} for details). It is called the $BFCG$ theory \cite{GirelliPfeifferPopescu2008,FariaMartinsMikovic2011}, and has the action
\begin{equation}
S_{BFCG} = \int_{\cM} \langle B \wedge F \rangle_{\fg} + \langle C \wedge G \rangle_{\fh}\,.
\end{equation}
Here the second term consists of a $\fh$-valued one-form Lagrange multiplier $C$, and a curvature three-form $G = \rmd \beta + A\wedge \beta$ for the $\fh$-valued two-form $\beta$, where $\fh$ is the Lie algebra of the group $\cH$. The pair $(A,\beta)$ is called the $2$-connection of the given $2$-group, while the pair $(F,G)$ is the corresponding $2$-curvature. The $\langle\;\,,\;\rangle_{\fh}$ is the invariant nondegenerate symmetric bilinear form in $\fh$, which is $\fg$-invariant.

The importance of the $BFCG$ theory for the Poincar\'e $2$-group, defined by the choice $\cG = SO(3,1)$ and $\cH=\realni^4$, lies in the fact that one can construct the action for GR by simply adding an additional term to the $BFCG$ action, called the simplicity constraint:
\begin{equation} \label{SpincubeDejstvo}
S = \int_{\cM} \langle B \wedge R \rangle_{\fg} + \langle e \wedge G \rangle_{\fh} - \langle \phi \wedge \left(B - \hodge (e\wedge e) \right) \rangle_{\fg} \, .
\end{equation}
Here we have relabeled $C\equiv e$ and $F\equiv R$, since in the case of the Poincar\'e $2$-group these fields have the interpretation of the tetrad field and the curvature two-form for the spin connection $A\equiv \omega$. The $\fg$-valued two-form $\phi$ is an additional Lagrange multiplier, featuring in the simplicity constraint term. The $\hodge$ is the Hodge dual operator for the Minkowski space. See \cite{TwoPoincare} for details.

The construction of (\ref{SpincubeDejstvo}) is in full analogy to the Plebanski model, where GR is constructed by adding a suitable simplicity constraint to the $BF$ theory based on the Lorentz group. However, in contrast to the Plebanski model, the constrained $BFCG$ model has one big advantage. Namely, the Lagrange multiplier $C$ has the interpretation of the tetrad field, which is therefore explicitly present in the topological sector of the action. This is not the case for the Plebanski model, where the simplicity constraint merely infers the implicit existence of the tetrad fields. Upon the covariant quantization, the Plebanski action gives rise to spin-foam models, while the constrained $BFCG$ action gives rise to the spin-cube model. Then, the explicit presence of the tetrad in (\ref{SpincubeDejstvo}) enables us to easily couple matter fields to gravity in the spin-cube model, in contrast to spin-foam models where this is a notoriously hard problem.

As a classical theory, the constrained $BFCG$ action lends itself also to the canonical quantization programme. In the canonical approach, the first and crucial step is to perform the Hamiltonian analysis of the theory, study the algebra of constraints, and eliminate second class constraints from the theory. However, due to the technical complexity of the Hamiltonian analysis, it is wise to discuss the pure $BFCG$ theory first, leaving the constrained theory for later. That is the aim of this paper. We will perform the full Hamiltonian analysis of the unconstrained $BFCG$ theory based on the Poincar\'e $2$-group, as a preparation for the more complicated case of the constrained $BFCG$ theory (\ref{SpincubeDejstvo}). The similar analysis has been done for the $BF$ theory in \cite{BFtheory}, and our analysis represents the generalization of that work to the $BFCG$ case.

We should note that the Hamiltonian analysis of the $BFCG$ model has already been done in a gauge-fixed form in \cite{MikovicOliveira2014}. In this paper we improve those results by providing a gauge-invariant canonical analysis. 

There is an interesting relationship between the $BFCG$ theory for the Poincar\'e $2$-group on one hand, and the topological Poincar\'e gauge theory on the other. Perhaps surprisingly, the two theories are equivalent, while their Hamiltonian structure is vastly different. This was discussed to an extent in \cite{MikovicOliveira2014}, but the full Hamiltonian analysis presented here lends itself nicely to a more complete comparison of the Hamiltonian formulations for the $BFCG$ theory and the topological Poincar\'e gauge theory. The results of this comparison are especially intriguing, and provide additional insight into the structure of the theory.

The paper is organized as follows. In section \ref{SecII} we introduce in detail the actions for the $BFCG$ theory and the topological Poincar\'e gauge theory, give a short overview of the Lagrange equations of motion, and prepare for the Hamiltonian analysis. The bulk of the Hamiltonian analysis is done in section \ref{SecIII}. We evaluate the conjugate momenta for the fields, obtain the primary constraints and construct the Hamiltonian of the theory. Then we impose consistency conditions on the primary constraints, which leads to secondary constraints and some determined Lagrange multipliers. The consistency conditions of the secondary constraints turn out to be satisfied identically, and do not introduce any new constraints. The constraints are then separated into first and second class, and their algebra is computed. Finally, the number of physical degrees of freedom is calculated, and ends up being zero, confirming that the theory is indeed topological. Building on these results, in section \ref{SecIV} we construct the Dirac brackets, which facilitate the elimination of the second class constraints from the theory, leading to the reduction of the phase space. Section \ref{SecV} is devoted to the study of the properties of the reduced phase space, with the emphasis on the differences between the $BFCG$ model and the topological Poincar\'e gauge theory. Section \ref{SecVI} contains the discussion of the results and our concluding remarks. The Appendix provides some technical details about the derivation and the discussion of the Bianchi identities, used in the main text.

Our notation is as follows. The spacetime indices are denoted with lowercase Greek alphabet letters $\mu,\nu,\rho,\dots$ and take the values $0,1,2,3$. When discussing the foliation of spacetime into space and time, the spacetime indices are split as $\mu = (0,i)$, where the lowercase indices from the middle of the Latin alphabet $i,j,k,\dots$ take only spacelike values $1,2,3$. The Poincar\'e group indices are denoted with lowercase letters from the beginning of the Latin alphabet, $a,b,c,\dots$ and take the values $0,1,2,3$. They are raised and lowered with the Minkowski metric $\eta_{ab} = \diag (-1,1,1,1)$. Capital Latin indices $A,B,C,\dots$ represent multi-index notation, and are used to count the second class constraints. Antisymmetrization is denoted with the square brackets around the indices,
\begin{equation}
A_{[ab]} = \frac{1}{2}\left( A_{ab} - A_{ba} \right)\,.
\end{equation}
In order to simplify the notation involving Poisson brackets, we will adopt the following convention. The left quantity in every Poisson bracket is assumed to be evaluated at the point $x=(t,\bi{x})$, while the right quantity at the point $x'=(t,\bi{x}')$. In addition, we use the shorthand notation for the 3-dimensional Dirac delta function $\delta^{(3)}\equiv \delta^{(3)}(\bi{x}-\bi{x}')$. This allows us to write an explicit but bulky expression like
\begin{equation}
\pb{A(t,\bi{x})}{B(t,\bi{x}')} = C(t,\bi{x}) \delta^{(3)}(\bi{x}-\bi{x}')
\end{equation}
more compactly as
\begin{equation}
\pb{A}{B} = C \delta^{(3)}\,,
\end{equation}
without ambiguity. This notation will be used systematically unless explicitly stated otherwise.

\section{\label{SecII}BFCG action}

The $BFCG$ theory for the Poincar\'e $2$-group is defined by the action
\begin{equation} \label{BFCGdejstvo}
S_{BFCG} = \int_{\cM} B_{ab} \wedge R^{ab} + e^a \wedge G_a \,.
\end{equation}
The variables of this action are the one-forms $e^a$, $\omega^{ab}$ and the two-forms $B^{ab}$, $\beta^a$. The curvatures $R^{ab}$ and $G^a$ are the field strengths of the 2-connection $(\omega^{ab}, \beta^a)$,
\begin{equation} \label{DefinicijaKrivine}
R^{ab} = \rmd \omega^{ab} + \omega^a{}_{c} \wedge \omega^{cb}\,,
\end{equation}
\begin{equation} \label{DefinicijaGkrivine}
G^a = \nabla \beta^a \equiv \rmd \beta^a + \omega^a{}_b \wedge \beta^b\,.
\end{equation}
The fields $B^{ab}$ and $e^a$ play the role of the Lagrange multipliers. Usually one would denote the latter multiplier as $C^a$, but we shall instead label it as $e^a$ since it will be interpreted as the tetrad field. Similarly, the usual notation for the connection one-form and its field strength is $A$ and $F$ respectively, but in our case they are denoted $\omega^{ab}$ and $R^{ab}$, since they are interpreted as the spin connection and the Riemann curvature two-form.

It is also convenient to introduce torsion as the field strength for the tetrad $e^a$,
\begin{equation} \label{DefinicijaTorzije}
T^a = \nabla e^a \equiv \rmd e^a + \omega^a{}_b \wedge e^b\,.
\end{equation}
Then, performing a partial integration in the second term in (\ref{BFCGdejstvo}) and using the Stokes theorem one can rewrite the action as
\begin{equation} \label{PGTdejstvo}
S_{TPGT} = \int_{\cM}  B_{ab} \wedge R^{ab} + \beta^a \wedge T_a - \int_{\del\cM} e^a\wedge \beta_a \,,
\end{equation}
where now $B^{ab}$ and $\beta^a$ play the role of Lagrange multipliers. Aside from the immaterial boundary term, this action represents the topological Poincar\'e gauge theory (TPGT). In order to fully appreciate the relationship between the two theories in the sense of the Hamiltonian analysis, let us introduce a parameter $\xi\in\realni$ and rewrite the action as
\begin{equation} \label{NaseDejstvo}
S = \int_{\cM} B_{ab} \wedge R^{ab} + \xi e^a \wedge G_a + (1-\xi) \beta^a \wedge T_a \,,
\end{equation}
where we have dropped the boundary term. It is obvious that the action (\ref{NaseDejstvo}) is a convenient interpolation between (\ref{BFCGdejstvo}) and (\ref{PGTdejstvo}), to which it reduces for the choices $\xi=1$ and $\xi=0$, respectively. The action (\ref{NaseDejstvo}) will therefore be the starting point for the Hamiltonian analysis.

It is also clear that all three actions (\ref{BFCGdejstvo}), (\ref{PGTdejstvo}) and (\ref{NaseDejstvo}) give rise to the same set of equations of motion, since these do not depend on the boundary. Taking the variation of (\ref{NaseDejstvo}) with respect to all the variables, one obtains
\begin{equation}
\begin{array}{ccl}
\delta B:      & & R^{ab} = 0\,, \\
\delta \beta:  & & T^a = 0 \,, \\
\delta e:      & & G^a = 0\,, \\
\delta \omega: & & \nabla B^{ab} - e^{[a}\wedge \beta^{b]} = 0\, , \\
\end{array}
\end{equation}
where
\begin{equation}
\nabla B^{ab} \equiv \rmd B^{ab} + \omega^a{}_c \wedge B^{cb} + \omega^b{}_c \wedge B^{ac}\,.
\end{equation}
The first two equations of motion tell us that spacetime has Minkowski geometry, while the third and fourth determine $\beta^a$ and $B^{ab}$. As we shall see later, there are no local propagating degrees of freedom in the theory.

Finally, for the convenience of the Hamiltonian analysis, we need to rewrite both the action and the equations of motion in a local coordinate frame. Choosing $\rmd x^{\mu}$ as basis one-forms, we can expand the fields in the standard fashion:
\begin{equation} \label{KomponenteTetradeIkoneksije}
e^a = e^a{}_{\mu}\rmd x^{\mu}\,, \qquad
\omega^{ab} = \omega^{ab}{}_{\mu} \rmd x^{\mu}\,,
\end{equation}
\begin{equation} \label{KomponenteBiBeta}
B^{ab} = \frac{1}{2} B^{ab}{}_{\mu\nu} \rmd x^{\mu} \wedge \rmd x^{\nu}\,, \qquad
\beta^a = \frac{1}{2}\beta^a{}_{\mu\nu} \rmd x^{\mu} \wedge \rmd x^{\nu}\,.
\end{equation}
Similarly, the field strengths for $\omega$, $e$ and $\beta$ are
\begin{equation} \label{KomponenteSvihKrivina}
\begin{array}{ccl}
R^{ab} & = & \ds \frac{1}{2} R^{ab}{}_{\mu\nu} \rmd x^{\mu} \wedge \rmd x^{\nu}\,, \vphantom{\ds\int}\\
T^a & = & \ds \frac{1}{2} T^a{}_{\mu\nu} \rmd x^{\mu} \wedge \rmd x^{\nu}\,, \vphantom{\ds\int}\\
G^a & = & \ds \frac{1}{6} G^a{}_{\mu\nu\rho} \rmd x^{\mu} \wedge \rmd x^{\nu} \wedge \rmd x^{\rho}\,. \vphantom{\ds\int}\\
\end{array}
\end{equation}
Using the relations (\ref{DefinicijaKrivine}), (\ref{DefinicijaGkrivine}) and  (\ref{DefinicijaTorzije}), we can write the component equations
\begin{equation} \label{KomponenteJacinaPolja}
\fl\begin{array}{ccl}
R^{ab}{}_{\mu\nu} & = & \del_{\mu} \omega^{ab}{}_{\nu} - \del_{\nu} \omega^{ab}{}_{\mu} + \omega^a{}_{c\mu} \omega^{cb}{}_{\nu} - \omega^a{}_{c\nu} \omega^{cb}{}_{\mu}\,, \\
T^a{}_{\mu\nu} & = & \del_{\mu} e^a{}_{\nu} - \del_{\nu} e^a{}_{\mu} + \omega^a{}_{b\mu} e^b{}_{\nu} - \omega^a{}_{b\nu} e^b{}_{\mu}\,, \vphantom{\ds\int} \\
G^a{}_{\mu\nu\rho} & = & \del_{\mu} \beta^a{}_{\nu\rho} + \del_{\nu} \beta^a{}_{\rho\mu} + \del_{\rho} \beta^a{}_{\mu\nu}
+ \omega^a{}_{b\mu} \beta^b{}_{\nu\rho} + \omega^a{}_{b\nu} \beta^b{}_{\rho\mu} + \omega^a{}_{b\rho} \beta^b{}_{\mu\nu}\,. \\
\end{array}
\end{equation}

Substituting expansions (\ref{KomponenteTetradeIkoneksije}), (\ref{KomponenteBiBeta}) and (\ref{KomponenteSvihKrivina}) into the action, we obtain
\begin{equation}
\fl S = \int_{\cM} \rmd^4x\, \lc^{\mu\nu\rho\sigma} \left[ \frac{1}{4} B_{ab\mu\nu} R^{ab}{}_{\rho\sigma} +
\frac{\xi}{6} e_{a\mu} G^a{}_{\nu\rho\sigma} + \frac{1-\xi}{4} \beta_{a\mu\nu} T^a{}_{\rho\sigma} \right]\,.
\end{equation}
Assuming that the spacetime manifold has the topology $\cM = \Sigma\times\realni$, where $\Sigma$ is a 3-dimensional spacelike hypersurface, from the above action we can read off the Lagrangian, which is the integral of the Lagrangian density over the hypersurface $\Sigma$:
\begin{equation} \label{Lagranzijan}
\fl L = \int_{\Sigma} \rmd^3x\, \lc^{\mu\nu\rho\sigma} \left[ \frac{1}{4} B_{ab\mu\nu} R^{ab}{}_{\rho\sigma} +
\frac{\xi}{6} e_{a\mu} G^a{}_{\nu\rho\sigma} + \frac{1-\xi}{4} \beta_{a\mu\nu} T^a{}_{\rho\sigma} \right]\,.
\end{equation}
Finally, the component form of equations of motion is:
\begin{equation}
\begin{array}{l}
\ds R^{ab}{}_{\mu\nu} = 0\,, \qquad T^a{}_{\mu\nu} = 0\,, \qquad G^a{}_{\mu\nu\rho} = 0\,, \vphantom{\ds\int} \\
\ds \lc^{\lambda\mu\nu\rho} \left[ \nabla_{\rho} B^{ab}{}_{\mu\nu}  - e^{[a}{}_{\rho} \beta^{b]}{}_{\mu\nu}\right] = 0\,. \vphantom{\ds\int} \\
\end{array}
\end{equation}

\section{\label{SecIII}Hamiltonian analysis}

Now we turn to the Hamiltonian analysis of the $BFCG$ theory. A detailed review of the general formalism can be found for example in \cite{Blagojevic}, Chapter V. In addition, the equivalent procedure for the ordinary $BF$ theory has been done in \cite{BFtheory}. 

As a first step, we calculate the momenta $\pi$ corresponding to the field variables $B^{ab}{}_{\mu\nu}$, $e^a{}_{\mu}$, $\omega^{ab}{}_{\mu}$ and $\beta^a{}_{\mu\nu}$. Differentiating the Lagrangian with respect to the time derivative of the appropriate fields, we obtain the momenta as follows:
\begin{equation}
\begin{array}{lclcl}
\pi(B)_{ab}{}^{\mu\nu} & \raz = & \ds \frac{\delta L}{\delta \del_0 B^{ab}{}_{\mu\nu}} & = & 0\,, \\
\pi(e)_{a}{}^{\mu} & \raz = & \ds \frac{\delta L}{\delta \del_0 e^a{}_{\mu}} & = & \ds \frac{1-\xi}{2} \lc^{0\mu\nu\rho} \beta_{a\nu\rho}\,, \vphantom{\ds\int^A} \\
\pi(\omega)_{ab}{}^{\mu} & \raz = & \ds \frac{\delta L}{\delta \del_0 \omega^{ab}{}_{\mu}} & = & \ds \lc^{0\mu\nu\rho} B_{ab\nu\rho}\,, \vphantom{\ds\int^A} \\
\pi(\beta)_a{}^{\mu\nu} & \raz = &  \ds \frac{\delta L}{\delta \del_0 \beta^a{}_{\mu\nu}} & = & - \xi \lc^{0\mu\nu\rho} e_{a\rho}\,. \vphantom{\ds\int^A} \\
\end{array}
\end{equation}
None of the momenta can be solved for the corresponding ``velocities'', so they all give rise to primary constraints:
\begin{equation}
\begin{array}{lcl}
P(B)_{ab}{}^{\mu\nu} & \raz\equiv & \pi(B)_{ab}{}^{\mu\nu} \approx 0\,, \vphantom{\ds\frac{1}{2}} \\
P(e)_a{}^{\mu} & \raz\equiv & \ds \pi(e)_a{}^{\mu} - \frac{1-\xi}{2} \lc^{0\mu\nu\rho} \beta_{a\nu\rho} \approx 0\,, \vphantom{\ds\frac{1}{2}} \\
P(\omega)_{ab}{}^{\mu} & \raz\equiv & \pi(\omega)_{ab}{}^{\mu} - \lc^{0\mu\nu\rho} B_{ab\nu\rho} \approx 0\,, \vphantom{\ds\frac{1}{2}} \\
P(\beta)_a{}^{\mu\nu} & \raz\equiv & \pi(\beta)_a{}^{\mu\nu} + \xi \lc^{0\mu\nu\rho} e_{a\rho} \approx 0\,. \vphantom{\ds\frac{1}{2}} \\
\end{array}
\end{equation}
The weak, on-shell equality is denoted ``$\approx$'', as opposed to the strong, off-shell equality which is denoted by the usual symbol ``$=$''.

Next we introduce the fundamental simultaneous Poisson brackets between the fields and their conjugate momenta,
\begin{equation}
\begin{array}{lcl}
\pb{B^{ab}{}_{\mu\nu}}{\pi(B)_{cd}{}^{\rho\sigma}} & \raz = & 4 \delta^a_{[c} \delta^b_{d]} \delta^{\rho}_{[\mu} \delta^{\sigma}_{\nu]} \delta^{(3)}\,, \\
\pb{e^a{}_{\mu}}{\pi(e)_b{}^{\nu}} & \raz = & \delta^a_b \delta^{\nu}_{\mu} \delta^{(3)}\,, \\
\pb{\omega^{ab}{}_{\mu}}{\pi(\omega)_{cd}{}^{\nu}} & \raz = & 2 \delta^a_{[c} \delta^b_{d]} \delta^{\nu}_{\mu} \delta^{(3)}\,, \\
\pb{\beta^a{}_{\mu\nu}}{\pi(\beta)_b{}^{\rho\sigma}} & \raz = & 2 \delta^a_b \delta^{\rho}_{[\mu} \delta^{\sigma}_{\nu]} \delta^{(3)}\,, \\
\end{array}
\end{equation}
and we employ them to calculate the algebra of primary constraints,
\begin{equation} \label{AlgebraPrimarnihVeza}
\begin{array}{lcl}
\pb{P(B)^{abjk}}{P(\omega)_{cd}{}^i} & = & 4 \lc^{0ijk} \delta^a_{[c} \delta^b_{d]} \delta^{(3)}, \\
\pb{P(e)^{ak}}{P(\beta)_b{}^{ij}} & = & - \lc^{0ijk} \delta^a_b \delta^{(3)}, \\
\end{array}
\end{equation}
while all other Poisson brackets vanish. Note that the algebra of primary constraints is independent of $\xi$.

Next we construct the canonical, on-shell Hamiltonian:
\begin{equation}
\begin{array}{ccl}
H_c & = & \ds \int_{\Sigma} \rmd^3\bi{x} \left[ \frac{1}{4} \pi(B)_{ab}{}^{\mu\nu} \del_0 B^{ab}{}_{\mu\nu} + \pi(e)_a{}^{\mu} \del_0 e^a{}_{\mu} + \right. \\
 & & \ds \left. + \frac{1}{2} \pi(\omega)_{ab}{}^{\mu} \del_0 \omega^{ab}{}_{\mu} + \frac{1}{2} \pi(\beta)_a{}^{\mu\nu} \del_0 \beta^a{}_{\mu\nu} \right] -L \,. \\
\end{array}
\end{equation}
The factors $1/4$ and $1/2$ are introduced to prevent overcounting of variables. Using (\ref{KomponenteJacinaPolja}) and  (\ref{Lagranzijan}), one can re\-arrange the expressions such that all velocities are multiplied by primary constraints, and therefore vanish from the Hamiltonian. After some algebra, the resulting expression can be written as
\begin{equation}
\begin{array}{ccl}
H_c & = & \ds - \int \rmd^3\bi{x}\, \lc^{0ijk} \left[ \frac{1}{2} B_{ab0i} R^{ab}{}_{jk} + \frac{1}{6} e_{a0} G^a{}_{ijk} + \right. \\
 & &  \ds \left. + \frac{1}{2} \beta_{a0k} T^a{}_{ij} + \frac{1}{2} \omega_{ab0} \left( \nabla_i B^{ab}{}_{jk} - e^a{}_i \beta^b{}_{jk} \right) \right] \,, \\
\end{array}
\end{equation}
up to a boundary term. The canonical Hamiltonian does not depend on any momenta, but only on fields and their spatial derivatives. Also, note that it does not depend on $\xi$ either. Finally, introducing Lagrange multipliers $\lambda$ for each of the primary constraints, we construct the total, off-shell Hamiltonian:
\begin{equation} \label{TotalniHamiltonijan}
\begin{array}{ccl}
H_T & = & \ds H_c \!+\! \int \rmd^3\bi{x} \left[ \lambda(e)^a{}_{\mu} P(e)_a{}^{\mu} \!+\! \frac{1}{2} \lambda(\omega)^{ab}{}_{\mu} P(\omega)_{ab}{}^{\mu} \right. \\
 & & \ds \left. + \frac{1}{4} \lambda(B)^{ab}{}_{\mu\nu} P(B)_{ab}{}^{\mu\nu} + \frac{1}{2} \lambda(\beta)^a{}_{\mu\nu} P(\beta)_a{}^{\mu\nu} \right] \,. \\
\end{array}
\end{equation}

We proceed with the calculation of the consistency requirements for the primary constraints,
\begin{equation}
\dot{P} \equiv \pb{P}{H_T} \approx 0\,.
\end{equation}
Half of the consistency requirements will give the secondary constraints $S$, while the other half will determine some of the multipliers $\lambda$. In particular, requiring that
\begin{equation}
\begin{array}{lll}
\dot{P}(B)_{ab}{}^{0i} \approx 0\,, & & \dot{P}(e)_a{}^0 \approx 0\,, \\
\dot{P}(\beta)_a{}^{0i} \approx 0\,, & & \dot{P}(\omega)_{ab}{}^0 \approx 0\,, \\
\end{array}
\end{equation}
we obtain the following secondary constraints:
\begin{equation} \label{SekundarneVeze}
\begin{array}{lcl}
S(R)^{ab}{}_{jk} & \equiv & R^{ab}{}_{jk} \approx 0, \\
S(G)^a & \equiv & \lc^{0ijk}G^a{}_{ijk} \approx 0, \\
S(T)^a{}_{ij} & \equiv & T^a{}_{ij} \approx 0, \\
S(B)^{ab} & \equiv & \lc^{0ijk} \left( \nabla_i B^{ab}{}_{jk} - e^{[a}{}_i \beta^{b]}{}_{jk} \right) \approx 0. \\
\end{array}
\end{equation}
The remaining consistency conditions for the primary constraints,
\begin{equation}
\begin{array}{lll}
\dot{P}(B)_{ab}{}^{jk} \approx 0\,, & & \dot{P}(e)_a{}^k \approx 0\,, \\
\dot{P}(\beta)_a{}^{jk} \approx 0\,, & & \dot{P}(\omega)_{ab}{}^k \approx 0\,, \\
\end{array}
\end{equation}
determine the following multipliers:
\begin{equation}
\fl\begin{array}{lcl}
\lambda(\omega)^{ab}{}_i & \approx & \ds \nabla_i \omega^{ab}{}_0\,, \vphantom{\frac{1}{2}} \\
\lambda(\beta)^a{}_{ij} & \approx & \ds  2 \nabla_{[j} \beta^a{}_{i]0} - \omega^a{}_{b0} \beta^b{}_{ij} \,, \vphantom{\frac{1}{2}} \\
\lambda(e)^a{}_i & \approx & \ds \nabla_i e^a{}_0 -  \omega^a{}_{b0} e^b{}_i \,, \vphantom{\frac{1}{2}} \\
\lambda(B)^{ab}{}_{ij} & \approx & \ds 2 \nabla_{[j} B^{ab}{}_{i]0} + 2 \omega^{[a}{}_{c0} B^{b]c}{}_{ij} + 
 e^{[a}{}_0 \beta^{b]}{}_{ij} + e^{[a}{}_j \beta^{b]}{}_{0i} - e^{[a}{}_i \beta^{b]}{}_{0j} \,. \vphantom{\frac{1}{2}} \\
\end{array}
\end{equation}
This leaves the multipliers
\begin{equation}
\lambda(\omega)^{ab}{}_0\,, \quad
\lambda(\beta)^a{}_{0i}\,, \quad
\lambda(e)^a{}_0\,, \quad
\lambda(B)^{ab}{}_{0i}\,,
\end{equation}
undetermined.

As the next step we impose the consistency conditions for the secondary constraints (\ref{SekundarneVeze}),
\begin{equation}
\begin{array}{lll}
\dot{S}(R)^{ab}{}_{jk} \approx 0\,, & & \dot{S}(G)^a \approx 0\,, \\
\dot{S}(T)^a{}_{ij} \approx 0\,, & & \dot{S}(B)^{ab} \approx 0\,. \\
\end{array}
\end{equation}
After a straightforward but lengthy calculation, it turns out that all these conditions are identically satisfied, producing no new constraints and determining no additional multipliers. Therefore, at this point all the consistency conditions have been exhausted.

Once we have found all the constraints in the theory, we need to classify them into first and second class. While some of the second class constraints can already be read from (\ref{AlgebraPrimarnihVeza}), the classification is not easy since constraints are unique only up to linear combinations. The most efficient way to tabulate all first class constraints is to substitute all determined multipliers into the total Hamiltonian (\ref{TotalniHamiltonijan}) and rewrite it in the form
\begin{equation} \label{TotHamiltonijanKombVezaPrveKlase}
\fl\begin{array}{ccl}
H_T & = & \ds \int \rmd^3\bi{x} \left[ \frac{1}{2} \lambda(B)^{ab}{}_{0i} \,\phi(B)_{ab}{}^i + \lambda(e)^a{}_0 \,\phi(e)_a + \lambda(\beta)^a{}_i \,\phi(\beta)_a{}^i \right. \vphantom{\ds\int} \\
 & & \ds \hphantom{\int \rmd^3\bi{x}m} \left. + \frac{1}{2} \lambda(\omega)^{ab} \,\phi(\omega)_{ab} -\frac{1}{2} B_{ab0i} \,\phi(R)^{abi} - e_{a0} \,\phi(G)^a \right. \vphantom{\ds\int} \\
 & & \ds \hphantom{\int \rmd^3\bi{x}m} \left. - \beta_{a0i} \,\phi(T)^{ai} - \frac{1}{2}\omega_{ab0} \,\phi(\nabla B)^{ab} \right]\,. \vphantom{\ds\int} \\
\end{array}
\end{equation}
The quantities $\phi$ are linear combinations of constraints, but must all be first class, since the total Hamiltonian weakly commutes with all constraints. Written in terms of primary and secondary constraints, they are:
\begin{equation}
\begin{array}{lcl}
\phi(B)_{ab}{}^i & = & \ds P(B)_{ab}{}^{0i}\,, \vphantom{\ds\int} \\
\phi(e)_a & = & \ds P(e)_a{}^0\,, \vphantom{\ds\int} \\
\phi(\beta)_a{}^i & = & \ds P(\beta)_a{}^{0i}\,, \vphantom{\ds\int} \\
\phi(\omega)_{ab} & = & \ds P(\omega)_{ab}{}^0\,, \vphantom{\ds\int} \\
\phi(R)^{abi} & = & \ds \lc^{0ijk} S(R)^{ab}{}_{jk} - \nabla_j P(B)^{abij}\,, \vphantom{\ds\int} \\
\phi(G)^a & = & \ds \frac{1}{6} S(G)^a + \nabla_i P(e)^{ai} - \frac{1}{4} \beta_{bij} P(B)^{abij}\,, \vphantom{\ds\int} \\
\phi(T)^{ai} & = & \ds \frac{1}{2} \lc^{0ijk} S(T)^a{}_{jk} \!-\! \nabla_j P(\beta)^{aij} \!+\! \frac{1}{2} e_{bj} P(B)^{abij}, \vphantom{\ds\int} \\
\phi(\nabla B)^{ab}\!\! & = & \ds S(B)^{ab} + \nabla_i P(\omega)^{abi} - B^{[a}{}_{cij} P(B)^{b]cij} \vphantom{\ds\int} \\
 & & \ds - 2 e^{[a}{}_i P(e)^{b]i} - \beta^{[a}{}_{ij} P(\beta)^{b]ij} \,.  \\
\end{array}
\end{equation}
These are the first class constraints in the theory. The remaining constraints are second class:
\begin{equation}
\begin{array}{lll}
\chi(B)_{ab}{}^{jk}=P(B)_{ab}{}^{jk}\,, & & \chi(e)_a{}^i=P(e)_a{}^i\,, \vphantom{\ds\int} \\
\chi(\omega)_{ab}{}^i=P(\omega)_{ab}{}^i\,, & & \chi(\beta)_a{}^{ij}=P(\beta)_a{}^{ij}\,. \vphantom{\ds\int} \\ 
\end{array}
\end{equation}
In order to calculate the full algebra of constraints, it is convenient to express them as functions of fundamental variables, as follows:
\begin{equation} \label{VezePrveKlaseEksplicitno}
\fl\begin{array}{lcl}
\phi(B)_{ab}{}^i & = & \ds \pi(B)_{ab}{}^{0i}\,, \vphantom{\ds\int} \\
\phi(e)_a & = & \ds \pi(e)_a{}^0\,, \vphantom{\ds\int} \\
\phi(\beta)_a{}^i & = & \ds \pi(\beta)_a{}^{0i}\,, \vphantom{\ds\int} \\
\phi(\omega)_{ab} & = & \ds \pi(\omega)_{ab}{}^0\,, \vphantom{\ds\int} \\
\phi(R)^{abi} & = & \ds \lc^{0ijk} R^{ab}{}_{jk} - \nabla_j \pi(B)^{abij}\,, \vphantom{\ds\int} \\
\phi(G)^a & = & \ds \frac{\xi}{6} \lc^{0ijk} G^a{}_{ijk} + \nabla_i \pi(e)^{ai} - \frac{1}{4} \beta_{bij} \pi(B)^{abij}\,, \vphantom{\ds\int} \\
\phi(T)^{ai} & = & \ds \frac{1-\xi}{2} \lc^{0ijk} T^a{}_{jk} - \nabla_j \pi(\beta)^{aij} + \frac{1}{2} e_{bj} \pi(B)^{abij}, \vphantom{\ds\int} \\
\phi(\nabla B)^{ab} & = & \ds \nabla_i \pi(\omega)^{abi} - B^{[a}{}_{cij} \pi(B)^{b]cij} -
2 e^{[a}{}_i \pi(e)^{b]i} - \beta^{[a}{}_{ij} \pi(\beta)^{b]ij} \,, \vphantom{\ds\int} \\
\end{array}
\end{equation}
and
\begin{equation} \label{VezeDrugeKlaseEksplicitno}
\begin{array}{lcl}
\chi(B)_{ab}{}^{jk} & = & \ds \pi(B)_{ab}{}^{jk}\,, \vphantom{\ds\int} \\
\chi(e)_a{}^i & = & \ds \pi(e)_a{}^i -\frac{1-\xi}{2} \lc^{0ijk} \beta_{ajk} \,, \vphantom{\ds\int} \\
\chi(\omega)_{ab}{}^i & = & \ds \pi(\omega)_{ab}{}^i -\lc^{0ijk} B_{abjk} \,, \vphantom{\ds\int} \\
\chi(\beta)_a{}^{ij} & = & \ds \pi(\beta)_a{}^{ij} + \xi \lc^{0ijk} e_{ak} \,. \vphantom{\ds\int} \\
\end{array}
\end{equation}
The algebra between the first class constraints is then
\begin{equation} \label{AlgebraVezaPrveKlase}
\begin{array}{lcl}
\pb{\phi(G)^a}{\phi(T)^{bi}} & = & \ds - \phi(R)^{abi} \delta^{(3)} \,, \vphantom{\ds\int} \\
\pb{\phi(G)^a}{\phi(\nabla B)_{cd}} & = & \ds 2 \delta^a_{[c} \phi(G)_{d]} \delta^{(3)} \,, \vphantom{\ds\int} \\
\pb{\phi(T)^{ai}}{\phi(\nabla B)_{cd}} & = & \ds 2 \delta^a_{[c} \phi(T)_{d]}{}^i \delta^{(3)} \,, \vphantom{\ds\int} \\
\pb{\phi(R)^{abi}}{\phi(\nabla B)_{cd}} & = & \ds - 4 \delta^{[a}_{[c} \phi(R)^{b]}{}_{d]}{}^i \delta^{(3)} \,, \vphantom{\ds\int} \\
\pb{\phi(\nabla B)^{ab}}{\phi(\nabla B)_{cd}} & = & \ds - 4 \delta^{[a}_{[c} \phi(\nabla B)^{b]}{}_{d]} \delta^{(3)} \,, \vphantom{\ds\int} \\
\end{array}
\end{equation}
the algebra between the second class constraints is, according to (\ref{AlgebraPrimarnihVeza}),
\begin{equation} \label{AlgebraVezaDrugeKlase}
\begin{array}{lcl}
\pb{\chi(B)^{abjk}}{\chi(\omega)_{cd}{}^i} & = & 4 \lc^{0ijk} \delta^a_{[c} \delta^b_{d]} \delta^{(3)}\,, \vphantom{\ds\int} \\
\pb{\chi(e)^{ak}}{\chi(\beta)_b{}^{ij}} & = & - \lc^{0ijk} \delta^a_b \delta^{(3)}\,, \vphantom{\ds\int} \\
\end{array}
\end{equation}
while the algebra between the first and second class constraints is
\begin{equation} \label{AlgebraVezaMesoviteKlase}
\begin{array}{lcl}
\pb{\phi(R)^{abi}}{\chi(\omega)_{cd}{}^j} & = & 4\delta^{[a}_{[c} \chi(B)^{b]}{}_{d]}{}^{ij} \delta^{(3)} \,, \vphantom{\ds\int} \\
\pb{\phi(G)^a}{\chi(\omega)_{cd}{}^i} & = & 2 \delta^a_{[c} \chi(e)_{d]}{}^i \delta^{(3)} \,, \vphantom{\ds\int} \\
\pb{\phi(G)^a}{\chi(\beta)_c{}^{jk}} & = & - \frac{1}{2} \chi(B)^{a}{}_{c}{}^{jk} \delta^{(3)} \,, \vphantom{\ds\int} \\
\pb{\phi(T)^{ai}}{\chi(\omega)_{cd}{}^j} & = & - 2\delta^a_{[c} \chi(\beta)_{d]}{}^{ij} \delta^{(3)} \,, \vphantom{\ds\int} \\
\pb{\phi(T)^{ai}}{\chi(e)_b{}^j} & = & \frac{1}{2} \chi(B)^a{}_b{}^{ij} \delta^{(3)} \,, \vphantom{\ds\int} \\
\pb{\phi(\nabla B)^{ab}}{\chi(\omega)_{cd}{}^i} & = & 4\delta^{[a}_{[c} \chi(\omega)_{d]}{}^{b]}{}^i \delta^{(3)} \,, \vphantom{\ds\int} \\
\pb{\phi(\nabla B)^{ab}}{\chi(\beta)_c{}^{jk}} & = & -2 \delta^{[a}_c \chi(\beta)^{b]jk} \delta^{(3)} \,, \vphantom{\ds\int} \\
\pb{\phi(\nabla B)^{ab}}{\chi(e)_c{}^i} & = & -2 \delta^{[a}_c \chi(e)^{b]i} \delta^{(3)} \,, \vphantom{\ds\int} \\
\pb{\phi(\nabla B)^{ab}}{\chi(B)_{cd}{}^{jk}} & = & 4\delta^{[a}_{[c} \chi(B)_{d]}{}^{b]jk} \delta^{(3)} \,. \vphantom{\ds\int} \\
\end{array}
\end{equation}
All other Poisson brackets among $\phi$ and $\chi$ are zero.

We see that the algebra is closed, and all Poisson brackets involving $\phi$ constraints weakly vanish, confirming that all $\phi$ are indeed first class. Also, the Poisson brackets between $\chi$ constraints do not weakly vanish, confirming that $\chi$ are indeed second class. Finally, note that the structure constants do not depend on $\xi$, despite the fact that the constraints $\phi$ and $\chi$ do.

The last main step in the Hamiltonian analysis is the counting of the physical degrees of freedom. Given $N$ initial independent fields in the theory, the dimension of the full phase space is $2N$. From this one subtracts the total number $F$ of first class constraints, the total number $S$ of second class constraints, and the total number $F$ of gauge fixing conditions. The result is the dimension of the physical phase space, $2n$, where $n$ is the number of physical degrees of freedom. Thus we have the general formula (see for example \cite{Blagojevic}),
\begin{equation} \label{JnaZaBrojStepeniSlobode}
n = N - F - \frac{S}{2}\, .
\end{equation}
The number of independent field components for each of the fundamental fields is
$$
\begin{array}{|c|c|c|c|} \hline
\omega^{ab}{}_{\mu} & \beta^a{}_{\mu\nu} & e^a{}_{\mu} & B^{ab}{}_{\mu\nu} \\ \hline
24 & 24 & 16 & 36 \\ \hline
\end{array}
$$
which gives the total $N=100$. Similarly, the number of independent components for the second class constraints is
$$
\begin{array}{|c|c|c|c|} \hline
\chi(B)_{ab}{}^{jk} & \chi(e)_a{}^i & \chi(\omega)_{ab}{}^i & \chi(\beta)_a{}^{ij} \\ \hline
18 & 12 & 18 & 12 \\ \hline
\end{array}
$$
which gives the total $S=60$. Regarding the first class constraints, the situation is a little more complicated, due to the presence of Bianchi identities (see the Appendix). In particular, not all components of $\phi(R)^{abi}$ and $\phi(T)^{ai}$ are independent. To see this, take the derivative of $\phi(R)^{abi}$ to obtain
\begin{equation}
\nabla_i \phi(R)^{abi} = \lc^{0ijk} \nabla_i R^{ab}{}_{jk} + R^{c[a}{}_{ij} \pi(B)_c{}^{b]ij}\, .
\end{equation}
The first term on the right-hand side is zero off-shell as a consequence of the second Bianchi identity (\ref{DrugiBI}). The second term on the right-hand side is also zero off-shell, since it is a product of two constraints,
\begin{equation}
R^{c[a}{}_{ij} \pi(B)_c{}^{b]ij} \equiv S(R)^{c[a}{}_{ij} P(B)_c{}^{b]ij} = 0\,.
\end{equation}
Therefore, we have the off-shell identity
\begin{equation} \label{OffShellIdentitetZaR}
\nabla_i \phi(R)^{abi} = 0\,,
\end{equation}
which means that $6$ components of $\phi(R)^{abi}$ are not independent of the others. In a similar fashion, we can calculate the following linear combination:
\begin{equation}
\begin{array}{l}
\ds \nabla_i \phi(T)^{ai} - \frac{1}{2}e_{bi} \phi(R)^{abi} = \vphantom{\ds\int} \hphantom{mmmmmmmmmmmm}\\
\hphantom{mmmmmm} \ds = \frac{1-\xi}{2}\lc^{0ijk} \left[ \nabla_i T^a{}_{jk} - R^{ab}{}_{ij} e_{bk} \right] \vphantom{\ds\int} \hphantom{mm} \\
\hphantom{mmmmmm} \ds - \frac{1}{2}S(R)^{ac}{}_{ij} \chi(B)_c{}^{ij} + \frac{1}{4} S(T)_{bij} P(B)^{abij}\,. \vphantom{\ds\int} \\
\end{array}
\end{equation}
The term in the square brackets is zero off-shell as a consequence of the first Bianchi identity (\ref{PrviBI}). Additionally, the remaining two terms are products of constraints, and therefore also zero off-shell. Thus we have another off-shell identity,
\begin{equation} \label{OffShellIdentitetZaT}
\nabla_i \phi(T)^{ai} - \frac{1}{2}e_{bi} \phi(R)^{abi} = 0\,,
\end{equation}
which means that $4$ components of $\phi(T)^{ai}$ are not independent of the others.

Taking (\ref{OffShellIdentitetZaR}) and (\ref{OffShellIdentitetZaT}) into account, the number of independent components of the first class constraints is
$$
\begin{array}{|c|c|c|c|c|c|c|c|} \hline
\phi(B)_{ab}{}^i & \phi(e)_a & \phi(\beta)_a{}^i & \phi(\omega)_{ab} & \phi(R)^{abi} & \phi(G)^a & \phi(T)^{ai} & \phi(\nabla B)^{ab} \\ \hline
18 & 4 & 12 & 6 & 18-6 & 4 & 12-4 & 6 \\ \hline
\end{array}
$$
which gives the total of $F=70$. Finally, substituting $N$, $F$ and $S$ into (\ref{JnaZaBrojStepeniSlobode}), we obtain:
\begin{equation} \label{BrojFizickihStepeniSlobode}
n = 100 - 70 - \frac{60}{2} = 0\,.
\end{equation}
We conclude that the theory has no physical degrees of freedom.

As the final point of the analysis, we note that one can introduce the following canonical transformation on the phase space of the theory:
\begin{equation} \label{VelikaKanonskaTransformacija}
\begin{array}{ccl}
\pi(\beta)_a{}^{ij} & \to & \ds \tilde{\pi}(\beta)_a{}^{ij} = \pi(\beta)_a{}^{ij} + \left( 1-2\xi \right) \lc^{0ijk} e_{ak}\,, \vphantom{\ds\int} \\
\pi(e)_a{}^i & \to & \ds \tilde{\pi}(e)_a{}^i = \pi(e)_a{}^i + \left( \frac{1}{2} - \xi \right) \lc^{0ijk} \beta_{ajk} \,, \vphantom{\ds\int} \\
\end{array}
\end{equation}
while all other fields and momenta map identically onto themselves. It is easy to check that this change of variables is indeed canonical, since it does not change the Poisson structure. Moreover, the Hamiltonian (\ref{TotHamiltonijanKombVezaPrveKlase}) and the primary and secondary constraints (\ref{VezePrveKlaseEksplicitno}) and (\ref{VezeDrugeKlaseEksplicitno}) all transform such that
\begin{equation}
\xi \to \tilde{\xi} = 1-\xi\,.
\end{equation}
This is a symmetry of the action (\ref{NaseDejstvo}) up to the boundary term, since
\begin{equation} \label{KsiSimetrijaDejstva}
S[1-\xi] = S[\xi] - \int_{\del\cM} e^a \wedge \beta_a\,.
\end{equation}
At the level of the full phase space, the canonical transformation (\ref{VelikaKanonskaTransformacija}) therefore maps between $\xi$ and $1-\xi$, in particular between the $BFCG$ theory ($\xi=1$) and the TPGT theory ($\xi=0$). Nevertheless, after the elimination of the second class constraints and the phase space reduction, the situation will be more complicated, as we shall see in section \ref{SecV}. The canonical transformation between the $BFCG$ and TPGT will still exist, but it will be singular in a certain sense, and not expressible in the generic form (\ref{VelikaKanonskaTransformacija}). This will be discussed in detail in section \ref{SecV}.

\section{\label{SecIV}Dirac brackets}

After the Hamiltonian analysis has been completed, we proceed to eliminate the second class constraints from the theory. This is done by introducing the Dirac brackets, defined as:
\begin{equation} \label{DirakoveZagradeDefinicija}
\fl\begin{array}{l}
\ds\db{F(t,\bi{x})}{G(t,\bi{x}')} = \pb{F(t,\bi{x})}{G(t,\bi{x}')} - \vphantom{\ds\int} \\
\hphantom{mmm} \ds - \int_{\Sigma} \rmd^3\bi{y} \int_{\Sigma} \rmd^3\bi{y}' \pb{F(t,\bi{x})}{\chi^A(t,\bi{y})} \Delta^{-1}_{AB}(t,\bi{y},\bi{y}') \pb{\chi^B(t,\bi{y}')}{G(t,\bi{x}')}\,,
\end{array}
\end{equation}
where $F$ and $G$ are some functions of the phase space variables, while the kernel $\Delta^{-1}_{AB}(t,\bi{y},\bi{y}')$ is the inverse of 
\begin{equation}
\Delta^{AB} (t,\bi{y},\bi{y}') \equiv \pb{\chi^A(t,\bi{y})}{\chi^B(t,\bi{y}')}\,.
\end{equation}
The multi-indices $A$ and $B$ count all $60$ independent second class constraints.

In order to evaluate the kernel $\Delta^{-1}$ and make the general definition (\ref{DirakoveZagradeDefinicija}) more manageable, we proceed in several steps. First, from the Poisson brackets (\ref{AlgebraVezaDrugeKlase}) we see that $\Delta^{AB}(t,\bi{x},\bi{y})$ is diagonal in the space variables $\bi{x}$ and $\bi{y}$, i.e. it can be written as
\begin{equation}
\Delta^{AB}(t,\bi{x},\bi{y}) = \Delta^{AB}(t) \delta^{(3)}(\bi{x}-\bi{y})\,.
\end{equation}
That means that its inverse will also be diagonal in those variables,
\begin{equation} \label{DijagonalnaInverznaDeltaMatrica}
\Delta^{-1}_{AB}(t,\bi{y},\bi{y}') = \Delta^{-1}_{AB} (t) \delta^{(3)}(\bi{y} - \bi{y}')\,,
\end{equation}
so that
\begin{equation}
\int_{\Sigma} \rmd^3{\bi{y}} \;\Delta^{AB}(t,\bi{x},\bi{y}) \Delta^{-1}_{BC}(t,\bi{y},\bi{y'}) = \delta^A_C \delta^{(3)}(\bi{x}-\bi{y}')\,,
\end{equation}
provided that
\begin{equation} \label{DefInverzneDeltaMatrice}
\Delta^{AB}(t) \Delta^{-1}_{BC}(t) = \delta^A_C\,.
\end{equation}
From now on we will drop the explicit dependence of time from the notation of these matrices, for convenience. Substituting (\ref{DijagonalnaInverznaDeltaMatrica}) into (\ref{DirakoveZagradeDefinicija}) and integrating over $\bi{y}'$, the Dirac brackets can be written in a simpler form
\begin{equation} \label{JednostavnijeDirakoveZagrade}
\db{F}{G} = \pb{F}{G} - \int_{\Sigma} \rmd^3\bi{y} \pb{F}{\chi^A(\bi{y})} \Delta^{-1}_{AB} \pb{\chi^B(\bi{y})}{G}\,,
\end{equation}
where we have again simplified the notation by implicitly assuming appropriate spacetime dependence of variables $F$ and $G$.

As a second step, if we rewrite $\chi^A$ and $\chi^B$ as quadruples
\begin{equation}
\begin{array}{ccl}
\chi^A & = & \ds \left( \chi(B)^{abij},\chi(\omega)^{abi},\chi(e)^{ai}, \chi(\beta)^{aij} \right)\,, \\
\chi^B & = & \ds \left( \chi(B)^{cdmn},\chi(\omega)^{cdm},\chi(e)^{cm}, \chi(\beta)^{cmn} \right)\,, \\
\end{array}
\end{equation}
we can write the matrix $\Delta^{AB}$ in the block-diagonal form,
\begin{equation}
\Delta^{AB} = \left( 
\begin{array}{cc|cc}
 & \Delta^{abij|cdm} &  & \\
\Delta^{abi|cdmn} & & & \\ \hline
 & & & \Delta^{ai|cmn} \\
 & & \Delta^{aij|cm} & \\
\end{array}
 \right) \,,
\end{equation}
where we have used vertical bars to separate row from column indices, and the blank entries in the matrix are assumed to be zero by convention. According to (\ref{AlgebraVezaDrugeKlase}) we have
\begin{equation} \label{ElementiDeltaMatrice}
\begin{array}{lcl}
\Delta^{abij|cdm} & = & \ds 4\lc^{0mij} \eta^{a[c} \eta^{d]b}\,, \\
\Delta^{abi|cdmn} & = & \ds - 4\lc^{0imn} \eta^{a[c} \eta^{d]b}\,, \\
\Delta^{ai|cmn} & = & \ds -\lc^{0mni} \eta^{ac}\,, \\
\Delta^{aij|cm} & = & \ds \lc^{0ijm} \eta^{ac}\,. \\
\end{array}
\end{equation}
The inverse matrix $\Delta^{-1}_{AB}$ then has a similar form,
\begin{equation}
\Delta^{-1}_{AB} = \left( 
\begin{array}{cc|cc}
 & \Delta^{-1}_{abij|cdm} &  & \\
\Delta^{-1}_{abi|cdmn} & & & \\ \hline
 & & & \Delta^{-1}_{ai|cmn} \\
 & & \Delta^{-1}_{aij|cm} & \\
\end{array}
 \right) \,.
\end{equation}
Using this, from (\ref{DefInverzneDeltaMatrice}) one can obtain the equations
\begin{equation}
\begin{array}{lcl}
\ds \frac{1}{2} \Delta^{abij|cdm} \Delta^{-1}_{cdm|a'b'i'j'} & = & \ds 4 \delta^a_{[a'} \delta^b_{b']} \delta^i_{[i'}\delta^j_{j']} \,,\vphantom{\ds\int} \\
\ds \frac{1}{4} \Delta^{abi|cdmn} \Delta^{-1}_{cdmn|a'b'i'} & = & \ds 2 \delta^i_{i'} \delta^a_{[a'} \delta^b_{b']} \,,\vphantom{\ds\int} \\
\ds \frac{1}{2} \Delta^{ai|cmn} \Delta^{-1}_{cmn|a'i'} & = & \ds \delta^a_{a'} \delta^i_{i'} \,,\vphantom{\ds\int} \\
\ds \Delta^{aij|cm} \Delta^{-1}_{cm|a'i'j'} & = & \ds 2 \delta^a_{a'} \delta^i_{[i'}\delta^j_{j']} \,,\vphantom{\ds\int} \\
\end{array}
\end{equation}
and then using (\ref{ElementiDeltaMatrice}) one can solve them to obtain the components of the inverse matrix,
\begin{equation}
\begin{array}{lcl}
\Delta^{-1}_{abij|cdm} & = & \ds \lc_{0ijm} \eta_{a[c} \eta_{d]b} \,, \\
\Delta^{-1}_{abi|cdmn} & = & \ds - \lc_{0imn} \eta_{a[c} \eta_{d]b} \,, \\
\Delta^{-1}_{ai|cmn} & = & \ds - \lc_{0imn} \eta_{ac} \,, \\
\Delta^{-1}_{aij|cm} & = & \ds \lc_{0ijm} \eta_{ac} \,. \\
\end{array}
\end{equation}
Here we have defined $\lc_{0123} \equiv - \lc^{0123}$.

Finally, the third step is to substitute the matrix $\Delta^{-1}_{AB}$ into (\ref{JednostavnijeDirakoveZagrade}) in order to obtain an explicit expression for the Dirac brackets:
\begin{equation} \label{KonacneDirakoveZagrade}
\db{F}{G} = \pb{F}{G} - \frac{1}{2} \lc_{0ijk} \int_{\Sigma} \rmd^3\bi{y} \; K^{ijk}(F,G,\bi{y})\,,
\end{equation}
where the kernel $K^{ijk}(F,G,\bi{y})$ is
\begin{equation}
\begin{array}{ccl}
K^{ijk}(F,G,\bi{y}) & = & \ds \frac{1}{4} \pb{F}{\chi(B)^{abij}(\bi{y})} \pb{\chi(\omega)_{ab}{}^k(\bi{y})}{G} \vphantom{\ds\int} \\
 & & \ds - \frac{1}{4} \pb{F}{\chi(\omega)^{abi}(\bi{y})} \pb{\chi(B)_{ab}{}^{jk}(\bi{y})}{G} \vphantom{\ds\int} \\
 & & \ds - \pb{F}{\chi(e)^{ai}(\bi{y})} \pb{\chi(\beta)_a{}^{jk}(\bi{y})}{G} \vphantom{\ds\int} \\
 & & \ds + \pb{F}{\chi(\beta)^{aij}(\bi{y})} \pb{\chi(e)_a{}^k(\bi{y})}{G} \,. \vphantom{\ds\int} \\
\end{array}
\end{equation}

Having constructed the Dirac brackets, the next task is to express the constraint algebra in terms of them. This has two main consequences. The first is that the Dirac bracket between any quantity and any second class constraint is automatically zero, by construction. This is obvious from the definition (\ref{DirakoveZagradeDefinicija}). The second is that after passing from Poisson brackets to Dirac brackets, the second class constraints can be set equal to zero off-shell, giving rise to the reduction of the phase space to one of its hypersurfaces. We will now concentrate on the constraint algebra, while the reduction of the phase space will be discussed in detail in the next section.

Looking at the algebra of constraints, (\ref{AlgebraVezaPrveKlase}), (\ref{AlgebraVezaDrugeKlase}) and (\ref{AlgebraVezaMesoviteKlase}), we see that it has the following rough structure:
\begin{equation}
\begin{array}{lllll}
\pb{\phi}{\phi} = \phi\,, & & \pb{\chi}{\chi} = \Delta\,, & & \pb{\phi}{\chi} = \chi\,, \\
\pb{\phi}{\phi} = 0\,, & & \pb{\chi}{\chi} = 0\,, & & \pb{\phi}{\chi} = 0\,. \\
\end{array}
\end{equation}
Knowing that the Dirac brackets between an arbitrary quantity and a second class constraint is zero by construction, we can immediately conclude that
\begin{equation}
\db{\chi}{\chi} = 0\,, \qquad \db{\phi}{\chi} = 0\,,
\end{equation}
which leaves only $\db{\phi}{\phi}$ to be discussed. For this, a schematic calculation gives:
\begin{equation}
\begin{array}{ccl}
\db{\phi}{\phi} & = & \ds \pb{\phi}{\phi} - \int \pb{\phi}{\chi} \Delta^{-1} \pb{\chi}{\phi} \vphantom{\ds\int} \\
 & = & \ds \pb{\phi}{\phi} + \int \chi \Delta^{-1} \chi \vphantom{\ds\int} \\
 & = & \ds \pb{\phi}{\phi} \,, \vphantom{\ds\int} \\
\end{array}
\end{equation}
due to the fact that the product of two constraints is zero off-shell. This is actually a proof that the algebra of primary constraints does not change when we pass from Poisson brackets to Dirac brackets. Therefore, we have:
\begin{equation}
\begin{array}{lcl}
\db{\phi(G)^a}{\phi(T)^{bi}} & = & \ds - \phi(R)^{abi} \delta^{(3)} \,, \vphantom{\ds\int} \\
\db{\phi(G)^a}{\phi(\nabla B)_{cd}} & = & \ds 2 \delta^a_{[c} \phi(G)_{d]} \delta^{(3)} \,, \vphantom{\ds\int} \\
\db{\phi(T)^{ai}}{\phi(\nabla B)_{cd}} & = & \ds 2 \delta^a_{[c} \phi(T)_{d]}{}^i \delta^{(3)} \,, \vphantom{\ds\int} \\
\db{\phi(R)^{abi}}{\phi(\nabla B)_{cd}} & = & \ds - 4 \delta^{[a}_{[c} \phi(R)^{b]}{}_{d]}{}^i \delta^{(3)} \,, \vphantom{\ds\int} \\
\db{\phi(\nabla B)^{ab}}{\phi(\nabla B)_{cd}} & = & \ds - 4 \delta^{[a}_{[c} \phi(\nabla B)^{b]}{}_{d]} \delta^{(3)} \,, \vphantom{\ds\int} \\
\end{array}
\end{equation}
while all other $\db{\phi}{\phi}$ vanish.

As a final point, note also that for an arbitrary quantity $A$ we have:
\begin{equation}
\begin{array}{ccl}
\db{A}{H_T} & = & \ds \pb{A}{H_T} - \int \pb{A}{\chi} \Delta^{-1} \pb{\chi}{H_T} \vphantom{\ds\int} \\
 & = & \ds \pb{A}{H_T} - \int \pb{A}{\chi} \Delta^{-1} \pb{\chi}{\phi} \vphantom{\ds\int} \\
 & = & \ds \pb{A}{H_T} - \int \pb{A}{\chi} \Delta^{-1} \chi\,, \vphantom{\ds\int} \\
\end{array}
\end{equation}
where we have used the fact that the total Hamiltonian (\ref{TotHamiltonijanKombVezaPrveKlase}) is a linear combination of first class constraints. The result can be rewritten as
\begin{equation}
\dot{A} = \db{A}{H_T} + \int \pb{A}{\chi} \Delta^{-1} \chi\,,
\end{equation}
which becomes the standard-looking equation of motion
\begin{equation}
\dot{A} = \db{A}{H_T}
\end{equation}
when one reduces the phase space by promoting $\chi\approx 0$ to off-shell equalities $\chi=0$. This reduction is the subject of the next section.

\section{\label{SecV}Phase space reduction}

The purpose of introducing Dirac brackets is to remove the second class constraints from the theory. When we use exclusively Dirac brackets, no result depends on second class constraints, and we can project all phase space points to the hypersurface defined by strong equalities $\chi = 0$, reducing its dimension from $2N$ to $2N-S$, without changing any physical property of the theory, in particular without breaking its gauge symmetry. In our case, from (\ref{VezeDrugeKlaseEksplicitno}) we have the following off-shell equations:
\begin{equation} \label{JakeVezeDrugeKlase}
\begin{array}{lcl}
\pi(B)_{ab}{}^{jk} & = & 0\,, \vphantom{\ds\int} \\
\pi(\omega)_{ab}{}^i -\lc^{0ijk} B_{abjk} & = & 0 \,, \vphantom{\ds\int} \\
\ds \pi(e)_a{}^i -\frac{1-\xi}{2} \lc^{0ijk} \beta_{ajk} & = & 0 \,, \vphantom{\ds\int} \\
\pi(\beta)_a{}^{ij} + \xi \lc^{0ijk} e_{ak} & = & 0 \,. \vphantom{\ds\int} \\
\end{array}
\end{equation}
We will discuss these equations in two steps, first by analyzing the two equations independent of $\xi$, and then discussing the $\xi$-dependent equations, which are more complicated. The first two equations can be rewritten as:
\begin{equation}
\pi(B)_{ab}{}^{ij} = 0\,, \qquad B^{ab}{}_{ij} = - \frac{1}{2} \lc_{0ijk} \pi(\omega)^{abk}\,.
\end{equation}
Note that we have expressed two conjugate variables in terms of other variables in the phase space. This reduces the full phase space to a hypersurface defined by these equations, namely orthogonal to the directions of $\pi(B)_{ab}{}^{ij}$ and diagonal in the directions of corresponding $(B,\pi(\omega))$ planes. Given that we have eliminated $36$ phase space variables, the dimension of the hypersurface is $200-36=164$, since the dimension of the full phase space was $2N=200$. On this hypersurface the expressions for some of the first class constraints (\ref{VezePrveKlaseEksplicitno}) simplify. In particular, the first four constraints remain unaffected, while the final four constraints become:
\begin{equation}
\begin{array}{lcl}
\phi(R)^{abi} & = & \ds \lc^{0ijk} R^{ab}{}_{jk} \,, \vphantom{\ds\int} \\
\phi(G)^a & = & \ds \frac{\xi}{6} \lc^{0ijk} G^a{}_{ijk} + \nabla_i \pi(e)^{ai} \,, \vphantom{\ds\int} \\
\phi(T)^{ai} & = & \ds \frac{1-\xi}{2} \lc^{0ijk} T^a{}_{jk} - \nabla_j \pi(\beta)^{aij} \,, \vphantom{\ds\int} \\
\phi(\nabla B)^{ab} & = & \ds \nabla_i \pi(\omega)^{abi} - 2 e^{[a}{}_i \pi(e)^{b]i} - \beta^{[a}{}_{ij} \pi(\beta)^{b]ij} \,. \vphantom{\ds\int} \\
\end{array}
\end{equation}

Let us now turn to $\xi$-dependent equations in (\ref{JakeVezeDrugeKlase}). We immediately see that there are two mutually incompatible cases, namely $\xi\neq 0$ and $\xi\neq 1$. In the $\xi\neq 0$ case, we see that we can solve the equations for $e$ and $\pi(e)$,
\begin{equation} \label{EprekoBeta}
e^a{}_i = \frac{1}{2\xi} \lc_{0ijk} \pi(\beta)^{ajk}\,, \qquad \pi(e)_a{}^i = \frac{1-\xi}{2}\lc^{0ijk} \beta_{ajk}\,,
\end{equation}
again expressing two conjugate variables in terms of remaining ones. This reduces the dimension of the hypersurface even further, down to $164-24=140$. If one does not impose any gauge fixing in the theory, this is the minimal dimension of the hypersurface, since we have in total $S=60$ second class constraints. The final three first class constraints simplify even further, as follows:
\begin{equation} \label{VezePrveKlaseKadKsiNijeNula}
\begin{array}{lcl}
\phi(G)^a & = & \ds \frac{1}{6} \lc^{0ijk} G^a{}_{ijk} \,, \vphantom{\ds\int} \\
\phi(T)^{ai} & = & \ds - \frac{1}{\xi} \nabla_j \pi(\beta)^{aij} \,, \vphantom{\ds\int} \\
\phi(\nabla B)^{ab} & = & \ds \nabla_i \pi(\omega)^{abi} - \frac{1}{\xi} \beta^{[a}{}_{ij} \pi(\beta)^{b]ij} \,. \vphantom{\ds\int} \\
\end{array}
\end{equation}
We should stress that none of these equations make sense in the case $\xi=0$, i.e. for the topological Poincar\'e gauge theory (\ref{PGTdejstvo}), but are completely valid for the case $\xi=1$, which represents the $BFCG$ theory (\ref{BFCGdejstvo}).

Alternatively, in the $\xi\neq 1$ case, it is not a good idea to solve (\ref{JakeVezeDrugeKlase}) for $e, \pi(e)$, since this cannot be done if $\xi=0$. Instead, we can solve for $\beta$ and $\pi(\beta)$,
\begin{equation} \label{BetaPrekoE}
\beta^a{}_{ij} = \frac{1}{\xi-1} \lc_{0ijk} \pi(e)^{ak}\,, \qquad \pi(\beta)_a{}^{ij} = - \xi \lc^{0ijk} e_{ak}\,.
\end{equation}
This time the phase space reduces to another hypersurface, different from the previous one, but again of the same dimension $164-24=140$. The final three first class constraints simplify again, however not to (\ref{VezePrveKlaseKadKsiNijeNula}), but to:
\begin{equation} \label{VezePrveKlaseKadKsiNijeJedan}
\begin{array}{lcl}
\phi(G)^a & = & \ds \frac{1}{1-\xi} \nabla_i \pi(e)^{ai} \,, \vphantom{\ds\int} \\
\phi(T)^{ai} & = & \ds \frac{1}{2} \lc^{0ijk} T^a{}_{jk} \,, \vphantom{\ds\int} \\
\phi(\nabla B)^{ab} & = & \ds \nabla_i \pi(\omega)^{abi} - \frac{2}{1-\xi} e^{[a}{}_i \pi(e)^{b]i} \,. \vphantom{\ds\int} \\
\end{array}
\end{equation}
In this case the choice $\xi=1$ does not make sense, which means that the $BFCG$ theory case is excluded. Nevertheless, the case $\xi=0$ is included, describing topological Poincar\'e gauge theory.

It is interesting to ask what happens in the case of generic $\xi$, when it is neither zero nor one. In that case one can solve (\ref{JakeVezeDrugeKlase}) either for $(e,\pi(e))$ or for $(\beta,\pi(\beta))$. The constraints can be expressed in either form (\ref{VezePrveKlaseKadKsiNijeNula}) or (\ref{VezePrveKlaseKadKsiNijeJedan}). It is important to note, though, that the resulting hypersurface depends on the choice of $\xi$. In this case one can calculate the Dirac brackets
\begin{equation} \label{DirakovaZagradaZaEiPiE}
\db{e^a{}_i}{\pi(e)_b{}^j} = (1-\xi) \delta^a_b \delta^j_i \delta^{(3)}
\end{equation}
and
\begin{equation} \label{DirakovaZagradaZaBetaIpiBeta}
\db{\beta^a{}_{ij}}{\pi(\beta)_b{}^{mn}} = 2\xi \delta^a_b \delta^m_{[i} \delta^n_{j]} \delta^{(3)}\,.
\end{equation}
It is then easy to verify that (\ref{BetaPrekoE}) is a canonical transformation from $(e,\pi(e))$ to $(\beta,\pi(\beta))$, with (\ref{EprekoBeta}) being the inverse transformation. In particular, as long as $\xi\neq 0,1$, this transformation maps (\ref{VezePrveKlaseKadKsiNijeJedan}) to (\ref{VezePrveKlaseKadKsiNijeNula}), and in addition maps (\ref{DirakovaZagradaZaEiPiE}) to (\ref{DirakovaZagradaZaBetaIpiBeta}), justifying its canonical nature.

However, the cases $\xi=0$ and $\xi=1$ are singular, and the canonical transformation (\ref{BetaPrekoE}), (\ref{EprekoBeta}) does not make sense for either of those. Nevertheless, there exists a singular canonical transformation which maps between those two cases (see \cite{MikovicOliveira2014}), given as:
\begin{equation} \label{SingularnaKanTransformacija}
\beta^a{}_{ij} = -\lc_{0ijk} \pi(e)^{ak}\,, \qquad \pi(\beta)_a{}^{ij} = -\lc^{0ijk} e_{ak}\,.
\end{equation}
In particular, for the case $\xi=0$ the Dirac bracket (\ref{DirakovaZagradaZaEiPiE}) evaluates to
\begin{equation} \label{KsiNulaDirakovaZagrada}
\db{e^a{}_i}{\pi(e)_b{}^j} = \delta^a_b \delta^j_i \delta^{(3)} \,,
\end{equation}
while the constraints (\ref{VezePrveKlaseKadKsiNijeJedan}) become
\begin{equation} \label{KsiNulaVeze}
\begin{array}{lcl}
\phi(G)^a & = & \ds \nabla_i \pi(e)^{ai} \,, \vphantom{\ds\int} \\
\phi(T)^{ai} & = & \ds \frac{1}{2} \lc^{0ijk} T^a{}_{jk} \,, \vphantom{\ds\int} \\
\phi(\nabla B)^{ab} & = & \ds \nabla_i \pi(\omega)^{abi} - 2 e^{[a}{}_i \pi(e)^{b]i} \,. \vphantom{\ds\int} \\
\end{array}
\end{equation}
On the other hand, when $\xi=1$ we have from (\ref{DirakovaZagradaZaBetaIpiBeta})
\begin{equation} \label{KsiJedanDirakovaZagrada}
\db{\beta^a{}_{ij}}{\pi(\beta)_b{}^{mn}} = 2 \delta^a_b \delta^m_{[i} \delta^n_{j]} \delta^{(3)} \,,
\end{equation}
and from (\ref{VezePrveKlaseKadKsiNijeNula}) we obtain
\begin{equation} \label{KsiJedanVeze}
\begin{array}{lcl}
\phi(G)^a & = & \ds \frac{1}{6} \lc^{0ijk} G^a{}_{ijk} \,, \vphantom{\ds\int} \\
\phi(T)^{ai} & = & \ds - \nabla_j \pi(\beta)^{aij} \,, \vphantom{\ds\int} \\
\phi(\nabla B)^{ab} & = & \ds \nabla_i \pi(\omega)^{abi} - \beta^{[a}{}_{ij} \pi(\beta)^{b]ij} \,. \vphantom{\ds\int} \\
\end{array}
\end{equation}
The transformation (\ref{SingularnaKanTransformacija}) then maps (\ref{KsiNulaDirakovaZagrada}) to (\ref{KsiJedanDirakovaZagrada}) and (\ref{KsiNulaVeze}) to (\ref{KsiJedanVeze}), with its inverse mapping everything the other way around.

To sum up, we have the following general situation. For a generic $\xi$ one can write the theory using either $(e,\pi(e))$ variables or $(\beta,\pi(\beta))$ variables, and there is a canonical transformation (\ref{BetaPrekoE}) connecting these two sets of variables, for the same value of $\xi$. For the singular cases $\xi=0$ and $\xi=1$ one does not have a choice which variables to use, but there exists the canonical transformation (\ref{SingularnaKanTransformacija}) which maps the $\xi=0$ theory into the $\xi=1$ theory, and vice versa. This canonical transformation is called singular because it cannot be obtained as a solution of the second class constraints (\ref{JakeVezeDrugeKlase}). In contrast to (\ref{BetaPrekoE}), which maps between various variables on the same hypersurface determined by the choice of $\xi$, the singular canonical transformation maps between two different hypersurfaces determined by choices $\xi=0$ and $\xi=1$. This establishes the relationship between the canonical structure of the $BFCG$ model and the canonical structure of the topological Poincar\'e gauge theory.

We should also discuss the status of the canonical transformation (\ref{VelikaKanonskaTransformacija}), which maps the full phase space onto itself, while inducing the transformation $\xi \to 1-\xi$. Note that (\ref{VelikaKanonskaTransformacija}) assumes that both momenta $\pi(e)_a{}^i$ and $\pi(\beta)_a{}^{ij}$ are variables in the phase space. However, after the reduction of the phase space using either (\ref{EprekoBeta}) or (\ref{BetaPrekoE}), one of the two momenta is eliminated, and is not a variable in the reduced phase space. Therefore, one cannot formulate the canonical transformation (\ref{VelikaKanonskaTransformacija}) as it stands, on the reduced phase space. Geometrically, every choice of $\xi$ specifies one particular reduced phase space, and the symmetry $\xi \to 1-\xi$ now maps between different spaces. While it is possible to construct a set of canonical transformations analogous to (\ref{VelikaKanonskaTransformacija}), in the sense of the map $\xi \to 1-\xi$, these are not derivable from (\ref{VelikaKanonskaTransformacija}). In particular, in the case of reduction (\ref{EprekoBeta}), the change of variables
\begin{equation} \label{KanTransfZaEprekoBeta}
\pi(\beta)_a{}^{ij} \to \tilde{\pi}(\beta)_a{}^{ij} = \frac{\xi}{1-\xi} \pi(\beta)_a{}^{ij}
\end{equation}
implements the map $\xi\to 1-\xi$ in the constraints (\ref{VezePrveKlaseKadKsiNijeNula}). Similarly, in the case of reduction (\ref{BetaPrekoE}), the change of variables
\begin{equation} \label{KanTransfZaBetaPrekoE}
\pi(e)_a{}^i \to \tilde{\pi}(e)_a{}^i = \frac{1-\xi}{\xi} \pi(e)_a{}^i\,,
\end{equation}
implements the same map in the constraints (\ref{VezePrveKlaseKadKsiNijeJedan}). Note that neither (\ref{KanTransfZaEprekoBeta}) nor (\ref{KanTransfZaBetaPrekoE}) can be derived from (\ref{VelikaKanonskaTransformacija}), and also that both transformations are defined only for $\xi\neq 0,1$. Finally, for the case $\xi=0 \to \xi=1$ and vice versa, we have the canonical transformation (\ref{SingularnaKanTransformacija}), which also cannot be inferred from (\ref{VelikaKanonskaTransformacija}). We can of course infer the existence of all these canonical transformations from the fact that the map $\xi \to 1-\xi$ is a symmetry of the theory, see (\ref{KsiSimetrijaDejstva}). But the actual forms of the transformations cannot be obtained from each other, due to the different sets of variables in respective phase spaces.

\section{\label{SecVI}Conclusions}

Let us summarize the results. We have studied the Hamiltonian structure of the $BFCG$ model (\ref{BFCGdejstvo}), and its relationship to the topological Poincar\'e gauge theory (\ref{PGTdejstvo}). In section \ref{SecII} we have defined both theories, proved the equivalence of their respective actions and Lagrange equations of motion, and introduced a generalized action (\ref{NaseDejstvo}) which depends on a real parameter $\xi$ and which reduces to the $BFCG$ theory for $\xi=1$, while it reduces to the topological Poincar\'e gauge theory for $\xi=0$. In this way, we could perform the Hamiltonian analysis of both theories simultaneously. This was done in section \ref{SecIII}, where it was established that there are no physical degrees of freedom, equation (\ref{BrojFizickihStepeniSlobode}), the algebra of constraints (\ref{AlgebraVezaPrveKlase}), (\ref{AlgebraVezaDrugeKlase}) and (\ref{AlgebraVezaMesoviteKlase}) has been computed, and the Hamiltonian of the theory written as a linear combination of first class constraints, equation (\ref{TotHamiltonijanKombVezaPrveKlase}). In section \ref{SecIV} we have constructed the Dirac brackets (\ref{KonacneDirakoveZagrade}), which facilitate the elimination of the second class constraints from the theory without breaking its gauge symmetry. The geometrical consequences of this were explored in section \ref{SecV}. The elimination of second class constraints projects the phase space onto one of its hypersurfaces, depending on the choice of the parameter $\xi$. Only at this point the difference between the $BFCG$ theory and the topological Poincar\'e gauge theory becomes observable, since they live on distinct hypersurfaces. For the generic hypersurface, when $\xi\neq 0,1$, the second class constraints can be solved in terms of different sets of variables, and for every choice of $\xi$ there is a canonical transformation which maps between these, mapping the hypersurface $\xi$ onto itself. However, this was not possible for the singular cases $\xi=0,1$. Instead, in these two cases there exists a singular canonical transformation which maps the $\xi=0$ hypersurface to $\xi=1$ hypersurface and vice versa, establishing the equivalence between the $BFCG$ theory and the topological Poincar\'e gauge theory, and clarifying the relationship between their respective variables.

The results obtained in this paper represent the straightforward generalization of results obtained in \cite{BFtheory} of the $BF$ theory based on the Lorentz group to the $BFCG$ theory based on the Poincar\'e $2$-group. Some of the material presented overlaps with \cite{MikovicOliveira2014}, but also improves on those previous results, in the following important ways. First, in this paper we have performed the full Hamiltonian analysis, as opposed to the shorthand procedure used in \cite{MikovicOliveira2014}. This facilitates a better basis for the future analysis of the constrained $BFCG$ theory (\ref{SpincubeDejstvo}), whose relevance is very high since it is equivalent to GR. Second, the procedure used in \cite{MikovicOliveira2014} was performed by employing a partial gauge fixing. This makes the calculations much simpler, but prevents us from computing the full algebra of constraints, in particular (\ref{AlgebraVezaDrugeKlase}) and (\ref{AlgebraVezaMesoviteKlase}). In this paper the gauge symmetry was kept intact, the full algebra of constraints has been computed, and proved to be closed. And third, in this paper we have given a more detailed analysis of the relationship between the $BFCG$ model and the topological Poincar\'e gauge theory, providing a better insight into the geometry of the reduced phase spaces for both theories.

The quantization of the theory has not been discussed, for two reasons. First, it was previously discussed in \cite{MikovicOliveira2014}. Second, while the formalism of Dirac brackets lends itself nicely to the canonical quantization, the quantum version of the $BFCG$ model itself is of limited interest. Instead, the relevant theory is the constrained $BFCG$ model (\ref{SpincubeDejstvo}), whose canonical quantization should give rise to a realistic theory of quantum gravity. This paper represents a first step toward the construction of such a theory, but the quantization procedure is however out of its scope, and will be studied in future work.

Another topic for future work would be the Hamiltonian analysis of the $BFCG$ model for an arbitrary $2$-group \cite{GirelliPfeifferPopescu2008,FariaMartinsMikovic2011}. On one hand, models based on the extensions of the Poincar\'e $2$-group can give rise to a theory of quantum gravity with matter fields, providing a basis for the study of unification of gravity with other interactions at the quantum level. On the other hand, various choices of $2$-groups can prove fruitful in the construction of new topological invariants of manifolds, which is a very important area of investigation in modern mathematics. For example, a state sum model based on the Euclidean $2$-group has been constructed in \cite{BaratinFreidel2015} (see also \cite{BaratinWise2009}). One could therefore study the Hamiltonian structure of that model and perform the canonical quantization. Its comparison to the state sum model should provide greater insight into the structure of the theory. Various other applications may also be possible.

\ack

MV would like to thank Prof. Milovan Vasili\'c for discussion and helpful suggestions. AM was partially supported by the FCT projects PEst-OE/MAT/UI0208/2013, EXCL/MAT-GEO/0222/2012, and by the bilateral project ``Quantum Gravity and Quantum Integrable Models - 2015-2016'' between Portugal and Serbia. MAO was supported by the FCT grant SFRH/BD/79285/2011. MV was supported by the FCT project PEst-OE/MAT/UI0208/2013, the FCT grant SFRH/BPD/46376/2008, the bilateral project ``Quantum Gravity and Quantum Integrable Models - 2015-2016'' between Portugal and Serbia, and partially by the project ON171031 of the Ministry of Education, Science and Technological Development, Serbia.

\appendix

\section*{Appendix. Bianchi identities}
\setcounter{section}{1}

Recalling the definitions of the torsion and curvature $2$-forms,
\begin{equation}
T^a = \rmd e^a + \omega^a{}_b \wedge e^b\,, \qquad R^{ab} = \rmd \omega^{ab} + \omega^a{}_c \wedge \omega^{cb}\,,
\end{equation}
one can take the exterior derivative of $T^a$ and $R^a$, and use the property $\rmd \rmd \equiv 0$ to obtain the following two identities:
\begin{equation}
\begin{array}{lcl}
\nabla T^a & \equiv & \rmd T^a + \omega^a{}_b \wedge T^b = R^a{}_b \wedge e^b\,, \\
\nabla R^{ab} & \equiv & \rmd R^{ab} + \omega^a{}_c \wedge R^{cb} + \omega^b{}_c \wedge R^{ac} = 0\,. \\
\end{array}
\end{equation}
These two identities are universally valid for torsion and curvature, and are called Bianchi identities. By expanding all quantities into components as
\begin{equation}
T^a = \frac{1}{2}T^a{}_{\mu\nu} \rmd x^{\mu} \wedge \rmd x^{\nu} \,, \qquad
R^{ab} = \frac{1}{2}R^{ab}{}_{\mu\nu} \rmd x^{\mu} \wedge \rmd x^{\nu} \,,
\end{equation}
\begin{equation}
e^a = e^a{}_{\mu} \rmd x^{\mu}\,, \qquad
\omega^{ab} = \omega^{ab}{}_{\mu} \rmd x^{\mu}\,,
\end{equation}
and using the formula $\rmd x^{\mu} \wedge \rmd x^{\nu} \wedge \rmd x^{\rho} \wedge \rmd x^{\sigma} = \lc^{\mu\nu\rho\sigma} \rmd^4x $, one can rewrite the Bianchi identities in component form as
\begin{equation}
\lc^{\lambda\mu\nu\rho} \left( \nabla_{\mu} T^a{}_{\nu\rho} - R^a{}_{b\mu\nu} e^b{}_{\rho} \right) = 0\,,
\end{equation}
and
\begin{equation}
\lc^{\lambda\mu\nu\rho} \nabla_{\mu} R^{ab}{}_{\nu\rho} = 0\,.
\end{equation}

For the purpose of Hamiltonian analysis, one can split the Bianchi identities into those which do not feature a time derivative and those that do. The time-independent pieces are obtained by taking $\lambda=0$ components:
\begin{equation} \label{PrviBI}
\lc^{0ijk} \left( \nabla_i T^a{}_{jk} - R^a{}_{bij} e^b{}_{k} \right) = 0\,,
\end{equation}
\begin{equation} \label{DrugiBI}
\lc^{0ijk} \nabla_i R^{ab}{}_{jk} = 0\,.
\end{equation}
These identities are valid as off-shell, strong equalities for every spacelike slice in spacetime, and can be enforced in all calculations involving the Hamiltonian analysis. The time-dependent pieces are obtained by taking $\lambda = i$ components:
\begin{equation}
\lc^{0ijk} \left( \nabla_0 T^a{}_{jk} - 2 \nabla_j T^a{}_{0k} - 2 R^a{}_{b0j} e^b{}_k - R^a{}_{bjk} e^b{}_0 \right) = 0\,,
\end{equation}
and
\begin{equation}
\lc^{0ijk} \left( \nabla_0 R^{ab}{}_{jk} - 2\nabla_j R^{ab}{}_{0k} \right) = 0\,.
\end{equation}
Due to the fact that they connect geometries of different spacelike slices in spacetime, they cannot be enforced off-shell. Instead, they can be derived from the Hamiltonian equations of motion of the theory.

In light of the Bianchi identities, we should note that the action (\ref{BFCGdejstvo}) features two more fields, $\beta^a$ and $B^{ab}$, which also have field strengths $G^a$ and $\nabla B^{ab}$, and for which one can similarly derive Bianchi-like identities,
\begin{equation}
\nabla G^a = R^a{}_b \wedge \beta^b, \qquad \nabla^2 B^{ab} = R^a{}_c \wedge B^{cb} + R^b{}_c \wedge B^{ac}\,.
\end{equation}
However, due to the fact that both $\beta^a$ and $B^{ab}$ are two-forms, in $4$-dimensional spacetime these identities will be single-component equations, with no free spacetime indices,
\begin{equation}
\lc^{\lambda\mu\nu\rho} \left( \frac{2}{3} \nabla_{\lambda} G^a{}_{\mu\nu\rho} - R^a{}_{b\mu\nu} \beta^b{}_{\nu\rho} \right) = 0\,,
\end{equation}
and similarly for $\nabla^2 B^{ab}$. Therefore, these equations necessarily feature time derivatives of the fields, and do not have a purely spatial counterpart to (\ref{PrviBI}) and (\ref{DrugiBI}). In this sense, like the time-dependent pieces of the Bianchi identities, they do not enforce any restrictions in the sense of the Hamiltonian analysis, but can instead be derived from the equations of motion and expressions for the Lagrange multipliers.

\end{document}